\documentclass[10pt,aps,twocolumn,floats,floatfix,amssymb,prd,superscriptaddress,nofootinbib]{revtex4-2}

\usepackage{graphicx}
\usepackage{amsmath,amssymb}
\usepackage{amsfonts}
\usepackage{xspace} 
\usepackage[usenames]{color}
\usepackage{dcolumn}
\usepackage{bm}
\usepackage{mathrsfs}
\usepackage[colorlinks=true]{hyperref}
\usepackage[all]{hypcap} 
\usepackage[utf8]{inputenc} 
\usepackage{multirow}
\usepackage{etoolbox}
\usepackage{tikz}
\usepackage[normalem]{ulem}

\newcommand{\G}{\mathcal{G}}
\newcommand{\R}{\mathfrak{R}}
\newcommand{\lag}{\mathcal{L}}

\newcommand{\D}{\mathrm{d}}

\newcommand{\intdx}{\int \D^4x \sqrt{-g}\,}

\newcommand{\be}{\begin{equation}}
\newcommand{\ee}{\end{equation}}
\newcommand{\bea}{\begin{eqnarray}}
\newcommand{\eea}{\end{eqnarray}}

\DeclareMathOperator{\Hessian}{Hess}

\begin{document}

%\title{Spontaneous Curvaturization of Black Holes}
\title{Purely metric Horndeski theories and spontaneous curvaturization of black holes}

\author{Astrid Eichhorn}
\email{eichhorn@thphys.uni-heidelberg.de}
\affiliation{Institut f\"ur Theoretische Physik, Universit\"at Heidelberg, Philosophenweg 16, 69120 Heidelberg, Germany}

\author{Pedro G. S. Fernandes}
\email{fernandes@thphys.uni-heidelberg.de}
\affiliation{Institut f\"ur Theoretische Physik, Universit\"at Heidelberg, Philosophenweg 16, 69120 Heidelberg, Germany}

\begin{abstract}
We explore purely metric theories of gravity with second-order equations of motion and a single additional, purely gravitational, propagating, scalar degree of freedom. We identify a subclass of these theories in which this scalar causes a phenomenon similar to black-hole scalarization, which we call \emph{curvaturization}: around small enough Kerr black holes, the scalar induces a tachyonic instability. This triggers a sudden growth of Ricci curvature and results in a new branch of vacuum black-hole solutions. We study the properties of these black holes both in the static as well as the spinning case.
\end{abstract}

\maketitle 

\section{Introduction}
According to General Relativity (GR), all vacuum black holes across the Universe ought to be described by the Kerr metric, and thus by just two parameters, their mass and their specific angular momentum. This result is known as black-hole uniqueness and it is a powerful assertion of GR, given that current observations of black holes span 9 orders of magnitude in mass \cite{KAGRA:2021duu,EventHorizonTelescope:2019dse,EventHorizonTelescope:2022wkp}. Few phenomena in nature are described by such simple models across such huge ranges in scales.

Beyond GR, the situation is entirely different. Already some of the simplest extensions of GR feature several branches of black-hole solutions, see, e.g., \cite{Lu:2015cqa,donevaScalarization2022,Burrage:2023zvk}. Understanding their properties and calculating their observational signatures in black-hole images and gravitational-wave signals has become a major research endeavour, see \cite{Ayzenberg:2023hfw,Barausse:2020rsu,ET:2019dnz} for reviews.
This endeavour is challenged by our very incomplete understanding of the space of modified theories of gravity. Starting out from Lovelock's theorem \cite{Lovelock:1971yv}, modifications of GR can be categorized and are based on i) additional degrees of freedom beyond GR, ii) non-localities, iii) extra dimensions. Thus, a burgeoning space of examples of modified theories is being developed, based on ad-hoc modifications of GR.

One popular example where black-hole uniqueness does not hold is given by scalar-tensor theories exhibiting \emph{black hole spontaneous scalarization} \cite{antoniouEvasionNoHairTheorems2018,bertiSpininducedBlackHole2021,cunhaSpontaneouslyScalarisedKerr2019,dimaSpininducedBlackHole2020,donevaNewGaussBonnetBlack2018,eichhornBreakingBlackholeUniqueness2023,herdeiroSpininducedScalarizedBlack2021,silvaSpontaneousScalarizationBlack2018,ventagliNeutronStarScalarization2021,Fernandes:2024ztk} (see Ref.~\cite{donevaScalarization2022} for a review) -- there are two classes of solutions that co-exist, one of them becoming dynamically preferred under certain conditions. These classes of solutions include a scalar-free Kerr black hole and a \emph{scalarized} solution. Notably, theories that exhibit scalarization deviate from GR only in the strong-field regime, making them modifications of gravity that remain consistent with current observational data \cite{donevaScalarization2022}.

The scalarization mechanism is driven by a scalar field. This scalar field is typically postulated in an ad-hoc fashion. More broadly, scalar fields are an ad-hoc addition in most theories in the Horndeski class, where the physical interpretation and origin of the additional scalar degree of freedom are usually unclear. One exception is given by the dilaton-Gauss-Bonnet-coupling which can be motivated from string-theoretic considerations \cite{Zwiebach:1985uq,Gross:1986mw}.

In this paper, we exploit field redefinitions to identify scalar-tensor theories that can be rewritten as purely geometric theories, i.e., with actions written entirely in terms of curvature invariants. This places these theories on an entirely different conceptual footing, because higher-order curvature terms are generic modifications of GR coming out of quantum-gravity theories. More specifically, higher-order curvature terms are generically expected to be part of the effective action that encodes the quantum dynamics of gravity below the Planck scale. Examples include higher-order curvature terms from string theory \cite{Zwiebach:1985uq,Gross:1986mw,Gross:1986iv,Grisaru:1986vi}, asymptotically safe gravity, e.g., \cite{Falls:2016wsa,Christiansen:2017bsy,Gubitosi:2018gsl}, reviewed in \cite{Eichhorn:2018yfc,Bonanno:2020bil,Knorr:2022dsx}, as well as \cite{Knorr:2018kog} for dynamical triangulations. Such higher-order curvature terms may give rise to a propagating scalar degree of freedom, e.g., the scalaron in $f(R)$ theories \cite{Sotiriou:2008rp,DeFelice:2010aj}. The scalaron emerges automatically from the action written in terms of the metric and can be thought of as a propagating physical scalar component of the metric. This is made explicit by a conformal transformation of the purely metric $f(R)$ theory, which isolates the dynamics for the scalar field and shows that it is the propagating conformal factor and not an ad-hoc addition to the degrees of freedom.
This is in stark contrast to many scalar-tensor theories, where an additional scalar is added in an ad-hoc fashion and need not have an interpretation as a scalar component of the metric.

In this paper, we pursue two goals: first, we identify such modified theories of gravity that propagate a single scalar degree of freedom that arises from the metric and is not added in an ad-hoc fashion. We  focus on such theories that admit a formulation in terms of second-order equations of motion. 
Second, we investigate the non-Kerr black-hole branch in these theories, expanding on the model presented in \cite{Liu:2020yqa}. Specifically, we derive and explore a subclass of $f(R, \mathcal{G})$ theories, where $\mathcal{G}$ is the Gauss-Bonnet invariant, featuring only one propagating scalar degree of freedom in addition to the metric. In this theory, the field equations are such that the Kerr metric remains a solution. As in $f(R)$ theories, the scalar degree of freedom is directly linked to spacetime curvature and is not introduced independently, but rather emerges naturally with a clear physical interpretation. This scalaron induces a tachyonic instability in Kerr black holes when they are sufficiently small, leading to their spontaneous \emph{``curvaturization"} -- a rapid growth in Ricci scalar curvature.

The paper is organized as follows. In Sec.~\ref{app:puregravity}, we explore the most general class of theories that extend the $f(R)$ framework while preserving the same number of degrees of freedom -- purely metric theories that can be reformulated as Horndeski theories. In Sec.~\ref{sec:curvaturization}, we begin by reviewing Starobinsky's gravity \cite{Starobinsky:1980te}, and extend it in such a way that the new theory exhibits spontaneous curvaturization and falls within the class discussed in Sec.~\ref{app:puregravity}. We present the field equations of the theory and show how vacuum GR solutions become tachyonically unstable. Then, in Sec.~\ref{sec:numerical}, we describe the numerical setup and boundary conditions used to derive static curvaturized black holes, followed by the presentation of our main numerical results -- a thorough investigation of the domain of existence of curvaturized black holes and their properties. In Sec.~\ref{sec:ricci-coupling}, we further extend the theory to include non-linear interactions associated with a Ricci-scalar-type coupling. These non-linear interactions significantly influence the properties of curvaturized black holes. In Sec.~\ref{sec:rotating} we generalize the static curvaturized solutions to rotating ones and explore their properties. We conclude in Sec.~\ref{sec:conclusions}.

\section{Purely Metric Horndeski Theories}
\label{app:puregravity}

In the space of effective actions -- or, equivalently, modified theories --  written purely in terms of curvature invariants, many theories are very likely not well-defined in that they do not admit a stable time evolution.
It is widely believed that Ostrogradsky's theorem provides a suitable criterion to rule out theories on the grounds of instability. Ostrogradsky's theorem was originally formulated in classical mechanics states that a theory that has a non-degenerate Lagrangian with higher than second order time derivatives, has a Hamiltonian that is not bounded from below. It is usually assumed that unboundedness of the Hamiltonian results in catastrophic instabilities.

However, the situation appears to be more subtle. First, the Ostrogradsky theorem can be evaded by introducing non-localities \cite{Barnaby:2007ve} or degeneracies \cite{Langlois:2015cwa}. Second, non-trivial counterexamples are known in classical mechanics, where an unbounded Hamiltonian does not lead to instabilities, even in a situation where the seemingly ``ghost" degrees of freedom do not decouple from the physical degrees of freedom \cite{Deffayet:2023wdg}. Third, even in gravity, the situation appears to be much more subtle than has long been expected. On the one hand, quadratic gravity, also known as Stelle gravity, has long been the ``poster child" for a theory plagued by ghost instabilities \cite{Stelle:1977ry}. However, numerical simulations call this expectation into question and yield stationary as well as dynamical spacetimes in which no exponentially fast runaway behavior occurs \cite{East:2023nsk,Held:2023aap}. Fourth, it has recently been proven that higher-order curvature terms, giving rise to higher-order time derivatives, can be rewritten as effective field theories with a well-posed initial value problem \cite{Figueras:2024bba}.

At present, no complete characterization of theories that have an unbounded Hamiltonian but nevertheless admit stable time evolution, exists. The space of viable modified theories of gravity is therefore likely larger than expected. Within this space, we focus on a subset of theories, by assuming that the restriction to second-order equations of motion, while \emph{not necessary}, may be \emph{sufficient} to guarantee that stable time evolution is possible. We therefore focus on Horndeski theories \cite{Horndeski:1974wa} and identify those that admit a formulation purely in terms of curvature invariants, without the ad-hoc addition of scalars. This naturally leads us to a subset of $f(R, \mathcal{G})$ theories, where $\G$ is the Gauss-Bonnet invariant
\begin{equation}
    \G = R^2 - 4 R_{\mu \nu}R^{\mu \nu} + R_{\mu \nu \alpha \beta} R^{\mu \nu \alpha \beta}.
\end{equation}
It has previously been observed that generic $f(R, \mathcal{G})$ theories feature instabilities about some physically interesting backgrounds, e.g., cosmological backgrounds that deviate from FLRW \cite{DeFelice:2010hg,DeFelice:2011ka,de2011linear}. These instabilities can be associated with one of the two propagating scalars that generic $f(R, \mathcal{G})$ theories have. Therefore, we are interested in $f(R, \mathcal{G})$ theories that only propagate a single scalar.

To derive the condition under which $f(R, \mathcal{G})$ theories contain only one dynamical scalar field, we introduce two auxiliary scalar fields, $\chi_1$ and $\chi_2$ such that
\begin{equation}
	\begin{aligned}
		&S = \intdx f(R,\G),\\&
		= \intdx \left[ (R-\chi_1)\partial_{\chi_1} f + (\G-\chi_2)\partial_{\chi_2} f + f(\chi_1,\chi_2) \right].
	\end{aligned}
\end{equation}
The field equations obtained by varying with respect to the scalar fields can be written in matrix form and are given by
\begin{gather}
	\begin{pmatrix} \partial_{\chi_1}^2f & \partial_{\chi_1 \chi_2}f \\ \partial_{\chi_1 \chi_2}f & \partial_{\chi_2}^2 f \end{pmatrix}
	\begin{pmatrix}
		R-\chi_1 \\
		\G-\chi_2
	\end{pmatrix}
	=
	 \begin{pmatrix}
	  0 \\
	  0
	  \end{pmatrix},
\end{gather}
where the matrix on the left-hand-side is the Hessian matrix of $f$, $\Hessian_f$. If the determinant of the Hessian matrix -- the Jacobian of the transformation $(R,\G) \to (\chi_1,\chi_2)$ -- is non-vanishing, there is a unique solution to the system given by $\chi_1 = R$ and $\chi_2 = \G$, and we can recover $\lag = f(R,\G)$ upon substituting the two scalar fields according to their equations of motion. 
If the determinant vanishes, the two scalar field equations are not independent,
\begin{equation}
	\det \Hessian_f \equiv (\partial_R^2 f)(\partial_\G^2 f) - (\partial_{R} \partial_\G f)^2 = 0.
	\label{eq:monge-ampere}
\end{equation}
Thus, we have introduced one too many scalars to account for the number of degrees of freedom of the theory, and the system is degenerate. 

The partial differential equation \eqref{eq:monge-ampere} is called \emph{homogeneous Monge-Ampére equation}, and there is no known general solution in terms of functions of the two variables $R$ and $\G$. Instead, the general solution to this partial differential equation is given in Ref.~\cite{HandbookPDEs} in parametric form. We (suggestively) denote this parameter $\R$, such that the parametric solution can be written as
\begin{eqnarray}
	%\begin{cases}
        f(R,\G) &=& \R R + \xi(\R) \G - V(\R),\label{eq:generalsolf}\\
        0&=&R + \xi'(\R) \G - V'(\R),  \label{eq:generalsolscalar}
    %\end{cases}
\end{eqnarray}
for arbitrary functions $\xi$ and $V$. The parameter $\R$ can be interpreted as a scalar field, and the functions $\xi$ and $V$ as a scalar-Gauss-Bonnet coupling and a scalar potential. The original $f(R, \G)$ theory can thus be written in the form
\begin{equation}
	S = \intdx \left( \R R + \xi(\R) \G - V(\R) \right).
	\label{eq:puregravity}
\end{equation}
The constraint \eqref{eq:generalsolscalar} can be interpreted as the scalar-field equation of motion, and for this reason no Lagrange multiplier is necessary to impose it. When the constraint is solved for $\R$ and substituted in the expression for $f(R,\G)$ in Eq.~\eqref{eq:generalsolscalar} we obtain a purely gravitational theory.

Given a degenerate $f(R,\G)$ theory, we can express it as a scalar-tensor theory \eqref{eq:puregravity} as follows. From Eq.~\eqref{eq:generalsolf}, we get the following set of equations
\begin{equation}
	\begin{cases}
        \partial_R f = \R,\\
        \partial_\G f = \xi(\R),\\
		R \partial_R f + \G \partial_\G f - f = V(\R).
    \end{cases}
    \label{eq:mapping}
\end{equation}
The first condition can be solved for $R$ or $\G$ as a function of $\R$, and substituted in the other two to get the functions $\xi(\R)$ and $V(\R)$. Interestingly, $V(\R)$ is the Legendre transform of $f(R,\G)$.\\

There is a second way of deriving the action Eq.~\eqref{eq:puregravity}, which starts from Horndeski gravity. It is illuminating to present this second derivation, because it makes explicit that the equations of motion of those $f(R, \G)$ theories that satisfy Eq.~\eqref{eq:monge-ampere} are second order.
The most general Lagrangian for a scalar-tensor theory with second-order equations of motion is the Horndeski Lagrangian \cite{Horndeski:1974wa}
\begin{equation}
    \begin{aligned}
		&\mathcal{L}_H = G_2(\R,X)-G_3(\R,X)\Box\R + G_4(\R,X)R
		\\&
        +G_{4X}  \left[(\Box\R)^2-\left(\nabla_\mu \nabla_\nu \R\right)^2\right]
          %\notag %\\ &
          +G_5(\R,X) G^{\mu\nu}\nabla_\mu \nabla_\nu \R
		  \\&
          -\frac{G_{5X}}{6}\left[
          (\Box\R)^3-3\Box\R \left(\nabla_\mu \nabla_\nu \R\right)^2
          +2\left(\nabla_\mu \nabla_\nu \R\right)^3
          \right],
    \end{aligned}
	\label{eq:Horndeski}
\end{equation}
where $\R$ is a scalar field and $X=-(\partial \R)^2/2$ is its kinetic term. 

To derive the most general Horndeski theory with a pure gravity formulation, we note the following: in the appropriate frame, the equation for the scalar field must not depend on derivatives of the scalar field. This ensures that the scalar field equation is a pure constraint and not a dynamical equation. Thus, it can be solved algebraically for the scalar field in terms of curvature invariant(s). Once this algebraic solution is obtained, it can be substituted back into the action to remove the scalar field, a situation that happens in both $f(R)$ and $f(\G)$ gravity. 
In what follows, it will be useful to know that a coupling $\xi(\R) \G$ is equivalent to taking the following combination of Horndeski functions \cite{kobayashiGeneralizedGinflationInflation2011,Kobayashi:2019hrl}
\begin{eqnarray}
        G_2 &=& 8 X^2(3-\log X) \xi'''',\nonumber\\
        G_3 &=& 4 X(7-3\log X) \xi''',\nonumber\\
        G_4 &=& 4 X(2-\log X) \xi'',\nonumber\\
        G_5 &=& -4 \left(\log X\right) \xi',\label{eq:HorndeskifG}
\end{eqnarray}
where the primes denote a derivative with respect to $\R$.
Consequently, $f(\G)$ theories are also part of the Horndeski class of theories, as is well-known \cite{Kobayashi:2019hrl}.

The most general scalar field equation of Horndeski's theory that does not involve derivatives of the scalar field can be obtained from the following Lagrangian, see \cite{Horndeski:1974wa}
\begin{eqnarray}
%	S = \intdx \left( \R R + \xi(\R) \G - V(\R) \right),
	G_2&=&-V(\R)+8 X^2(3-\log X) \xi'''',\nonumber\\
    G_3&=&4 X(7-3\log X) \xi''',\nonumber\\
    G_4&=& \R+4 X(2-\log X) \xi'',\nonumber\\
    G_5 &=& -4 \left(\log X\right) \xi',
    \label{eq:puregravityHorndeski}
\end{eqnarray}
where we have used Eq.~\eqref{eq:HorndeskifG}. This is the unique result up to scalar field redefinitions\footnote{A free function of the scalar field multiplying $R$ is allowed. However, upon a (possibly singular) field redefinition, we can bring the theory to this form.}, and where $\xi(\R)$ and $V(\R)$ are free functions of $\R$. Upon using that Eq.~\eqref{eq:HorndeskifG} combines to $\xi(\R)\G$, Eq.~\eqref{eq:puregravityHorndeski} gives rise to the same Lagrangian that defines the action \eqref{eq:puregravity}. We have therefore shown explicitly that this is the most general action that is on-shell equivalent to an $f(R, \G)$ action propagating only a single scalar field in addition to the metric and featuring second-order equations of motion.

In summary, we have an on-shell equivalence between a degenerate $f(R, \G)$ theory and a subset of Horndeski theories. This equivalence can be used in both directions: Starting from Horndeski theories, the equivalence singles out those theories in which there is no ad-hoc addition of a scalar field. Instead, the scalar has a \emph{geometric origin} in these theories. Conversely, starting from degenerate $f(R,\G)$ theories, the mapping to a scalar-tensor theories makes it easier to analyze phenomenological aspects of the theory, using insights obtained about scalar-tensor theories. One important example, that we expand upon in the sections that follow is black-hole scalarization: in addition to the Kerr branch, there is a second branch of vacuum black-hole solutions. In the frame in which the theory is written as a scalar-tensor theory, this branch features scalar hair, i.e., it is scalarized. Using the on-shell equivalence to $f(R, \G)$ theories, we recognize that the scalar hair is actually nothing but a nonzero configuration of the Ricci scalar, i.e., the black hole grows nontrivial ``curvature hair", i.e., it \emph{curvaturizes}.

We conclude our derivation of purely gravitational Horndeski theories by noting that we are not the first to study these theories. 
These degenerate theories were first noticed to behave differently in Ref.~\cite{DeFelice:2009ak} when studying perturbations on a Friedmann-Robertson-Walker background, and later shown to be in general healthier than generic $f(R,\G)$ theories \cite{DeFelice:2010hg,DeFelice:2011ka,de2011linear}. Ref.~\cite{DeFelice:2009ak} further argued that these theories should be equivalent to a scalar-tensor theory similar to Eq.~\eqref{eq:puregravity} given the interdependence of the two scalars. Ref.~\cite{Bueno:2016dol} showed that theories which are functions of the Lovelock invariants (in a generic number of spacetime dimensions) obeying a degeneracy condition analogous to Eq.~\eqref{eq:monge-ampere}  propagate a reduced number of degrees of freedom compared to general ones. Applying the results of Ref.~\cite{Bueno:2016dol} to the particular case of $f(R,\G)$ theories in four dimensions, we conclude that the degenerate theories obeying Eq.~\eqref{eq:monge-ampere} propagate only one additional scalar degree of freedom, instead of two.

\section{Spontaneous Curvaturization}
\label{sec:curvaturization}
In this section, we will find a general class of degenerate $f(R, \G)$ theories in which there is a second branch of black-hole solutions and the Kerr solution is not always stable. 
To do so, we use the equivalence to scalar-tensor theories, in which the conditions for black-hole scalarization are fairly well-understood. Due to the on-shell equivalence to a degenerate $f(R,\G)$ theory, the resulting black holes are better described as having undergone \emph{curvaturization}, i.e., the dynamical growth of a nonzero value of the Ricci scalar. At the same time, this puts these black-hole solutions on a distinct conceptual footing: whereas many deviations from Kerr black holes are proposed in the literature, the deviation from Kerr is often obtained by an ad-hoc addition of (non-minimally coupled) matter, or at the cost of introducing higher-order derivatives.
Here, the new branch arises purely geometrically, i.e., due to the presence of additional curvature invariants at the action.

Theories that exhibit spontaneous scalarization have to obey two conditions. First, spacetimes that are vacuum solutions of Einstein's equations must be vacuum solutions of the new theory with a trivial scalar field; second, the scalar must induce an instability around these GR backgrounds in certain situations.
Typically, this instability is tachyonic and is excited in the strong-field regime, e.g., around black holes. This is achieved by having a scalar field whose linear perturbations around a vacuum GR background obey a Klein-Gordon type equation with a squared effective mass
\begin{equation}
	\mu_{\rm eff}^2 = \mu^2 - \alpha \G,
	\label{eq:effmass}
\end{equation}
where $\alpha$ is a coupling constant with dimensions of length squared.\footnote{We work in units where $c=1=G$.} If the combination $\alpha \G$ achieves sufficiently negative values, a tachyonic instability occurs, and a new scalarized solution emerges.

To identify which degenerate $f(R,\G)$ theories scalarize, we start from Starobinsky gravity \cite{Starobinsky:1980te} and its scalar-tensor formulation.
The action for Starobinsky gravity is given by
\begin{equation}
	S = \intdx f(R), \quad \mathrm{with} \quad f(R) = R + \frac{R^2}{6\mu^2},
	\label{eq:starobinsky}
\end{equation}
where $\mu$ is a coupling constant that, as shown next, can be interpreted as the mass of a scalar.
This theory can be re-written as a scalar-tensor theory
\begin{equation}
	S = \intdx \left( \R R - \frac{3}{2} \mu^2 (\R-1)^2 \right),
	\label{eq:starobinskyST}
\end{equation}
where
\begin{equation}
    \R = \partial_R f = 1 + \frac{R}{3\mu^2}.
    \label{eq:scalaron1}
\end{equation}
The scalar is dynamical thanks to its coupling to the Ricci scalar and has a squared mass given by $\mu^2$. This becomes explicit by performing field redefinitions, namely a conformal transformation
\begin{equation}
	%g_{\mu \nu} = \frac{1}{\phi} \tilde{g}_{\mu \nu}, \quad \phi = e^{\sqrt{2/3}\varphi},
	\tilde{g}_{\mu \nu} = \R g_{\mu \nu}, \quad \varphi = \sqrt{\frac{3}{2}} \log \R,
	\label{eq:conformaltransf}
\end{equation}
such that in the new frame the new scalar is minimally coupled with a potential $\sim \mu^2 \varphi^2 + \mathcal{O}(\varphi^3)$ (see Appendix \ref{app:EinFrame} for details).

This is the point where we bridge the discussion on Starobinsky's $f(R)$ gravity and spontaneous scalarization. Noticing the role that the coupling constant of the $R^2$ term plays, namely that it is the mass of the propagating scalaron, we promote the scalaron's mass to be spacetime dependent as in Eq.~\eqref{eq:effmass}, $\mu^2 \to \mu^2 - \alpha \G$, in the action \eqref{eq:starobinsky} such that
\begin{equation}
    S = \intdx f(R,\G),
    \label{eq:scalarization1}
\end{equation}
with
\begin{equation}
    f(R,\G) = R + \frac{R^2}{6\left(\mu^2 - \alpha \G\right)}. 
	\label{eq:frg}
\end{equation}
This is a purely gravitational theory (i.e., only depends on the metric tensor $g_{\mu \nu}$) belonging to the well-known $f(R,\G)$ class and was first considered in Ref.~\cite{Liu:2020yqa} (see also Ref.~\cite{Liu:2024wvw}). Its Hessian matrix is degenerate, such that it is on-shell equivalent to a Horndeski theory.\footnote{The equations of motion derived from Eq.~\eqref{eq:frg} are actually fourth-order in derivatives. The on-shell equivalence to Horndeski gravity implies that the fourth order in derivatives do not give rise to ghost degrees of freedom. This is similar to what happens in $f(R)$ theories, which also have higher order equations of motion. However, the higher-order terms do not affect the spin-2-sector. Instead, the scalar, which is non-dynamical in GR, becomes dynamical, but is not a ghost.}
In particular, the action \eqref{eq:scalarization1} can be written as Eq.~\eqref{eq:puregravity} with
\begin{equation}
	\xi(\R) = \frac{3}{2} \alpha (\R-1)^2, \quad V(\R) = \frac{3}{2}\mu^2(\R-1)^2,
	\label{eq:couplings1}
\end{equation}
and where
\begin{equation}
    \R = \partial_R f = 1 + \frac{R}{3(\mu^2-\alpha \G)},
    \label{eq:scalaron2}
\end{equation}
is the scalaron, much like it was defined in the $f(R)$ case \eqref{eq:scalaron1}.
The equivalence between the actions \eqref{eq:scalarization1} and \eqref{eq:puregravity} becomes again evident when substituting the relation \eqref{eq:scalaron2} into Eq.~\eqref{eq:puregravity}, where the theory \eqref{eq:scalarization1} is recovered.

The field equations of the theory \eqref{eq:scalarization1} are obtained by varying with respect to the metric
\begin{equation}
	\begin{aligned}
		&\left( G_{\mu \nu} + g_{\mu \nu}\Box - \nabla_\mu \nabla_\nu \right)\R + \frac{1}{2} g_{\mu \nu} V(\R) \\&- 4 {}^* R^*_{\mu \alpha \nu \beta} \nabla^\alpha \nabla^\beta \xi(\R) = 0,
	\end{aligned}
    \label{eq:EinEqs}
\end{equation}
where ${}^* R^*_{\mu \alpha \nu \beta}$ is the double-dual of the Riemann tensor, and again $\R=\partial_R f$ is defined in Eq.~\eqref{eq:scalaron2}.\footnote{In the scalar-tensor picture where $\R$ is taken to be an independent field, its equation yields $R + \xi'(\R)\G - V'(\R) = 0$, which is equivalent to Eq.~\eqref{eq:scalaron2}.}
The trace of the field equations \eqref{eq:EinEqs} together with Eq. \eqref{eq:scalaron2} gives a propagation equation for $\R$
\begin{equation}
    \begin{aligned}
        \Box \R = \frac{1}{3} \bigg(& - \R \,\xi'(\R) \G + 4 G^{\mu \nu} \nabla_\mu \nabla_\nu \xi(\R)\\& + \R V'(\R) - 2 V(\R)\bigg),
    \end{aligned}
    \label{eq:SFeq2}
\end{equation}
where the new dynamical degree of freedom becomes manifest.

The Kerr metric is a solution of the theory \eqref{eq:scalarization1}. In this case the scalaron curvature scalar becomes $\R=1$ because $R = 0$. 

Next, we check whether the Kerr metric is prone to a tachyonic instability such that $\R \neq 1$ and thus $R\neq0$ is realized. To that end, we consider small perturbations of the system around these solutions, where the scalaron is perturbed as
\begin{equation}
	\R \approx 1 + \delta \R,
\end{equation}
and its linearized propagation equation becomes
\begin{equation}
    (\Box - \mu_{\rm eff}^2) \delta \R = 0.
	\label{eq:perttachyon}
\end{equation}
Herein, the squared effective mass $\mu_{\rm eff}^2$ is the one given in Eq.~\eqref{eq:effmass}. Therefore, we expect that Kerr black holes with small enough masses \cite{donevaNewGaussBonnetBlack2018} or high-enough spin \cite{dimaSpininducedBlackHole2020} become unstable and spontaneously evolve into a configuration with a non-zero Ricci curvature scalar -- in other words, black holes curvaturize. 

The Lagrangian density $f(R,\G)$ given in Eq.~\eqref{eq:frg} is not the most general one exhibiting curvaturization, and we can generalize the results in \cite{Liu:2020yqa}. For example, any Lagrangian of the form
\begin{equation}
    S = \intdx R\left(1 + \sum_{n=1}^{N_{\rm max}}c_n \left[ \frac{R}{6\left(\mu^2 - \alpha \mathcal{G}\right)}\right]^n\right),
    \label{eq:genact1}
\end{equation}
obeys the degeneracy condition \eqref{eq:monge-ampere} and exhibits curvaturization for constant couplings $c_n$, if $c_1 \neq 0$. Even if the sum is performed over real numbers $\geq 1$, instead of just integers, this statement remains true. 

Theories that exhibit scalarization face challenges regarding stability of the solutions as well as observational constraints, e.g., from binary pulsars. It has been shown that adding a non-minimal couplings of the scalar field to the Ricci scalar can address these challenges \cite{ventagliNeutronStarScalarization2021,antoniouBlackHoleScalarization2021,antoniouCompactObjectScalarization2021}, although some issues with instabilities may remain \cite{Kleihaus:2023zzs,Blazquez-Salcedo:2024rvb}. Therefore, we would like to understand whether this can be extended to theories in which black holes curvaturize. %Indeed, it turns out that, 
We find that similar to the extension by a Ricci coupling in the standard scalarization setting, where $\alpha \G \rightarrow \alpha \G + \beta R$, we can introduce a Ricci coupling here. 
Indeed, the Lagrangian
\begin{equation}
    f(R,\G) = R + \frac{R^2}{6\left(\mu^2 - \alpha \G + \beta R\right)},
    \label{eq:scalarization2}
\end{equation}
 contains the dimensionless Ricci coupling $\beta$ and falls into the class of models introduced in Eq.~\eqref{eq:genact1}. In fact, Eq.~\eqref{eq:scalarization2} corresponds to the sum in Eq.~\eqref{eq:genact1} with $c_n = (-6\beta)^{n-1}$ and an infinite number of terms, $N_{\rm max} \to \infty$. 

The theory can be re-cast as a scalar-tensor theory belonging to the Horndeski class, and it takes the form of Eq.~\eqref{eq:puregravity} with functions
\begin{equation}
	\begin{aligned}
		&\xi(\R) = \frac{\alpha \left(\sqrt{1-6\beta(\R-1)}-1\right)^2}{6\beta^2}, \\&
		V(\R) = \frac{\mu^2 \left(\sqrt{1-6\beta(\R-1)}-1\right)^2}{6\beta^2},
	\end{aligned}
	\label{eq:couplingsricci}
\end{equation}
which coincide with those of Eq.~\eqref{eq:couplings1} to zeroth order in $\beta$, but possess an infinite tower of higher-order corrections, introducing non-linearities into the system. 

Before we explore curvaturized black-hole solutions in these theories, let us make a few general comments.
The Lagrangians in Eq.~\eqref{eq:genact1} are nonlocal in that they depend on inverse powers of a curvature invariant and thus even the simplest representative of this family of actions already features an expansion in local curvature invariants to arbitrarily high powers. Nonlocal actions appear naturally as infrared endpoints of Renormalization Group flows, i.e., nonlocalities generically arise when quantum fluctuations of massless fields such as the graviton are integrated out, see, e.g., \cite{Codello:2015mba} and references therein, although these typically involve derivative operators. %Thus
Nevertheless, a nonlocal action of Eq.~\eqref{eq:genact1} may potentially arise as (part of) the effective action in an appropriate setting. It is in this spirit, as a phenomenological ansatz for the effective action, that we consider the non-local Lagrangians in this paper.

The occurrence of a nonlocality can also be made plausible by the following consideration: a generic $f(R, \mathcal{G})$ action propagates two scalar degrees of freedom. We aim at removing one of them. A well-known method to reduce the number of propagating degrees of freedom is to write non-local theories, see \cite{Barnaby:2007ve} for examples. Indeed, if we expand the Lagrangian in Eq.~\eqref{eq:frg} in $\alpha \G/\mu^2$ to finite order, the degeneracy condition that removes one propagating degree of freedom does not hold. Thus, the nonlocality is necessary to remove the extra degree of freedom, while at the same time enabling curvaturization.

\section{Curvaturized Black Holes}
\label{sec:numerical}
Curvaturization of static black holes occurs when the Schwarzschild solution is affected by %the 
a
tachyonic instability. 
For Schwarzschild black holes in the absence of a bare mass $\mu$ for the scalaron, the threshold of instability is known to occur when $M/\sqrt{\alpha} \lesssim 1.174$, if $\alpha>0$ \cite{donevaNewGaussBonnetBlack2018}, which we assume to be the case in the rest of the paper. When $\mu \neq 0$, this threshold shifts such that the instability occurs for smaller values of the black hole mass \cite{Macedo:2019sem}. In the rest of this work, when considering numerical solutions, we will focus our attention on the simplest case, namely we will assume that the mass of the scalar is negligible compared to the black-hole mass.\footnote{For $\mu \rightarrow 0$, the flat-space limit of the action Eq.~\eqref{eq:scalarization2} is subtle; we thus in principle keep $\mu \neq 0$, but focus on $\mu \ll M$ for black-hole solutions with ADM mass $M$.} Thus, we already expect that the new, curvaturized black-hole branch agrees with the Schwarzschild solution for $M/\sqrt{\alpha}=1.174$. Starting from this point, we investigate the domain of existence of the curvaturized solution numerically.

To obtain static and spherically symmetric black holes of the theory \eqref{eq:scalarization2} (both for $\beta=0$ and $\beta \neq 0$) we work with a line-element given by
\begin{equation}
	\D s^2 = - f(r) e^{-2\delta(r)} \D t^2 + \frac{\D r^2}{f(r)} + r^2 \left( \D \theta^2 + \sin^2\theta \D \Phi^2 \right).
\end{equation}
To simplify the problem, we consider the scalar-tensor formulation given in Eq.~\eqref{eq:puregravity} with $\R$ an independent field since the field equations reduce to second-order ones in all fields. This is beneficial from a numerical point of view. Since $f$ and $\delta$ depend only on the radial coordinate $r$, from Eq.~\eqref{eq:scalaron2} so does $\R$. The field equations for this line-element reduce to two first order differential equations for the metric functions $f$ and $\delta$, and one second-order equation for $\R$.

\subsection{Numerical Method}
We solve the system of three ordinary differential equations for $f$, $\delta$ and $\R$ using the Runge-Kutta 4(5) method implemented in \textsc{Julia}'s differential equations library \cite{DifferentialEquations.jl-2017}. We use the symmetry of the field equations under a global scaling symmetry (amounting to a rescaling of $M$) to fix the location of the horizon $r_H=2$; solutions with $r_H \neq 2$ can later be obtained by a global scaling. For each pair $(\alpha,\R_H)$, where $\R_H$ is the value of the scalar at the horizon, we integrate outwards starting from
a distance $r-r_H \sim 10^{-12}$ from the horizon towards large $r \sim 10^6$ using initial conditions obtained from the near-horizon expansion (see below).

To determine $\R_H$, we use that for each $\alpha$ not all values $\R_H$  yield the correct asymptotic behavior, namely $\lim_{r\to \infty} \R = 1$. For this reason we impose this condition by using a shooting method on $\R_H$. Moreover, we have independently implemented this problem in the code developed in Ref.~\cite{Fernandes:2022gde}, observing agreement between the two codes to high accuracy.

The initial conditions are obtained from the near-horizon expansion as follows:
Imposing the existence of a horizon at $r=r_H$ and regularity, we are led to the expansion
\begin{equation}
	\begin{aligned}
		& f = f'_H (r-r_H) + \mathcal{O}\left((r-r_H)^2\right),\\&
		\delta = \delta_H + \delta'_H (r-r_H) + \mathcal{O}\left((r-r_H)^2\right),\\&
		\R = \R_H + \R'_H (r-r_H) + \mathcal{O}\left((r-r_H)^2\right).
		\label{eq:horizonexp}
	\end{aligned}
\end{equation}
Solving the field equations order-by-order in powers of $r-r_H$, we get
\begin{equation}
	\begin{aligned}
		& f'_H = \frac{2}{r_H + \sqrt{r_H^2 - 8 \xi'(\R_H)}},\\&
		\delta'_H = \frac{2r_H}{3(r_H^2+4\xi'(\R_H))},\\&
		\R'_H = \R_H \frac{\sqrt{r_H^2-8\xi'(\R_H)}}{r_H^2 + 4 \xi'(\R_H)}.
		\label{eq:asympexp}
	\end{aligned}
\end{equation}
By requiring that the first derivatives of the metric functions at the horizon are real and non-divergent, we find conditions for the existence of solutions
\begin{equation}
	r_H^2 \geq 8\xi'(\R_H), \quad \mathrm{and} \quad r_H^2 + 4 \xi'(\R_H) \neq 0.
	\label{eq:existence_conditions}
\end{equation}

We obtain the asymptotic expansion by requiring that the spacetime is asymptotically flat. 
We expand the metric functions in powers of $1/r$ around Minkowski spacetime and solve the field equations order-by-order, obtaining
\begin{equation}
	\begin{aligned}
		& f = 1-\frac{2 M_{\rm ADM}}{r} + \mathcal{O}\left(1/r^2\right),\\&
		\delta = \frac{M_{\rm K}-M_{\rm ADM}}{r} + \mathcal{O}\left(1/r^2\right),\\&
		\R = 1 +\frac{M_{\rm K}-M_{\rm ADM}}{r} + \mathcal{O}\left(1/r^2\right),
	\end{aligned}
\end{equation}
where $M_{\rm K}$ and $M_{\rm ADM}$ are the Komar \cite{PhysRev.129.1873} and Arnowitt-Deser-Misner (ADM) \cite{PhysRev.122.997} masses, which do not coincide unless the solution is a Schwarzschild black hole.\footnote{The Komar and ADM masses can be extracted from the asymptotic decay of the $g_{tt}$ and $g_{rr}$ metric components, respectively.} This is due to the fact that we work in the Jordan frame, where there is a non-minimal coupling between the scalaron and the Ricci scalar.
The Einstein-frame metric $\tilde{g}_{\mu \nu} = \R g_{\mu \nu}$ has coinciding Komar and ADM masses. For this reason in what follows we take as our measure of mass the Einstein frame mass $M$, which coincides with the average of the Komar and ADM masses
\begin{equation}
	M = \frac{M_{\rm K} + M_{\rm ADM}}{2}.
\end{equation}
%Moreover, 
From the asymptotic decay of $\R$ we observe that the difference between the Komar and ADM masses 
(in the Jordan frame) can be interpreted as a charge of the scalaron
\begin{equation}
	Q = M_{\rm K}-M_{\rm ADM}.
\end{equation}
Note that in the Einstein frame, the scalar charge is also non-zero, but not expressed in terms of the difference of Komar and ADM masses, such that physical observables computed from it are frame-invariant.

\begin{figure*}[!t]
	\centering
	\includegraphics[width=\linewidth]{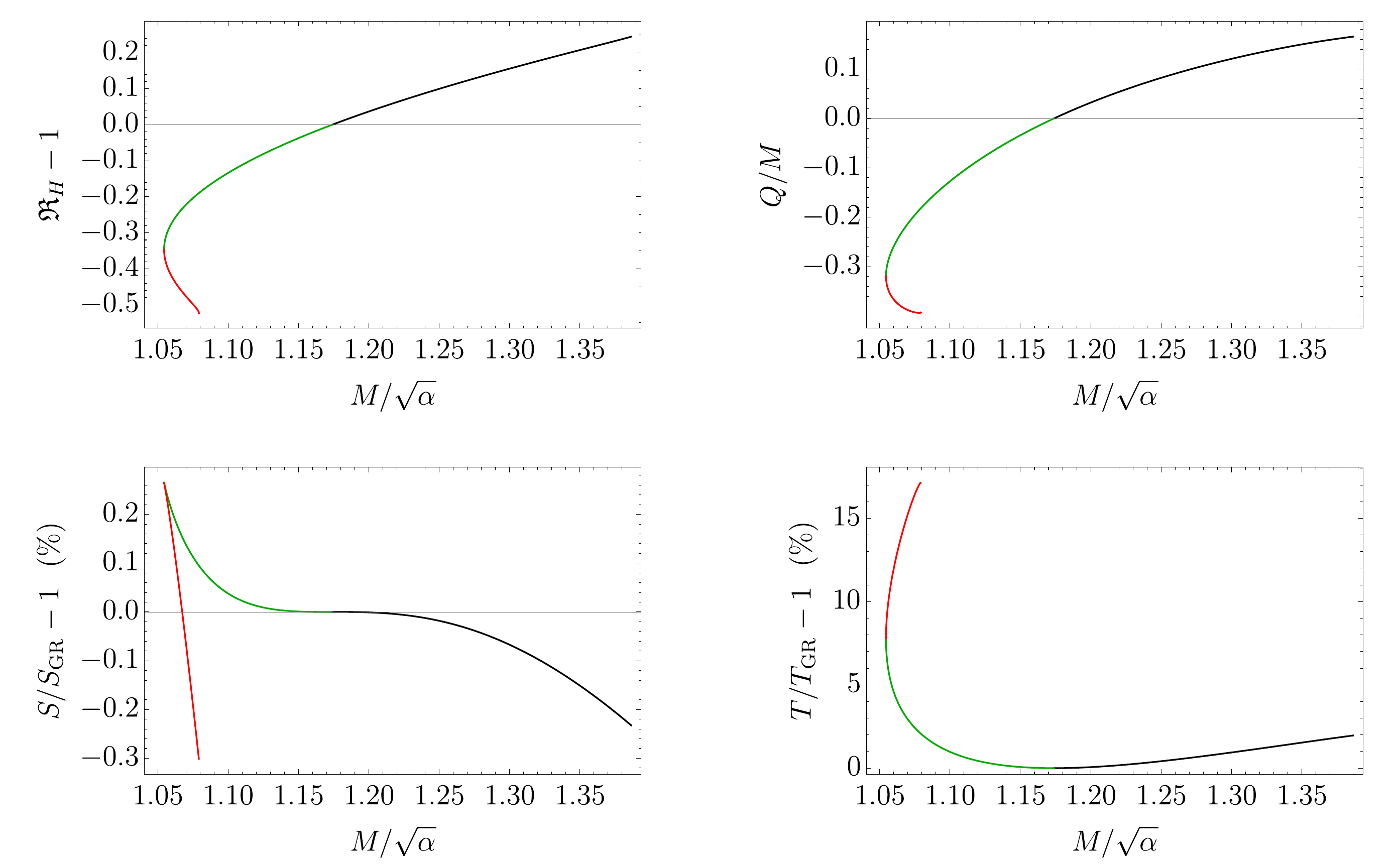}
    \caption{Domain of existence of curvaturized black holes and their physical quantities. The values of $\R_H$ (top-left), $Q$ (top-right), $S$ (bottom-left) and $T$ (bottom-right) of curvaturized black holes are presented as a function of $M/\sqrt{\alpha}$. The three distinct colors indicate a branch that is entropically disfavored (black line), entropically favored (green line, terminating in the minimum-mass-solution) and a branch on which entropic preference changes and which terminates in a solution that is singular on the horizon (red line).}
	\label{fig:plots}
\end{figure*}

\subsection{Curvaturized black holes at vanishing Ricci coupling}

In this section we perform a thorough analysis of curvaturized black holes, their domain of existence and properties, expanding on the analysis done in Ref.~\cite{Liu:2020yqa}.

In agreement with previous results in the literature \cite{donevaNewGaussBonnetBlack2018} and perturbation theory, we have found a new branch of black holes branching out of the GR solution for masses around $M/\sqrt{\alpha} \approx 1.174$. This can be observed in Fig.~\ref{fig:plots}, where we present the values of $\R_H$ (top-left), $Q$ (top-right), $S$ (bottom-left) and $T$ (bottom-right) of curvaturized black holes as a function of $M/\sqrt{\alpha}$.

The theory \eqref{eq:puregravity} is not symmetric around the vacuum expectation value of the scalaron, $\R=1$. Therefore, black hole solutions differ depending on whether $\R$ is greater or smaller than unity. This is shown in the top-left panel of Fig.~\ref{fig:plots}, where solutions with $\R_H>1$ exist for values $M/\sqrt{\alpha} > 1.174$ -- where the GR black holes are stable -- while solutions with $\R_H<1$ exist for values $M/\sqrt{\alpha} < 1.174$ -- where the GR black holes are unstable. We expect solutions with $\R_H>1$, where the curve tilts to the right, to be unstable/unphysical, much like the ones in the canonical scalarization scenario with a quadratic coupling \cite{silvaSpontaneousScalarizationBlack2018,Macedo:2019sem,antoniouStableSpontaneouslyscalarizedBlack2022}. This is because the endpoint of the tachyonic instability of the GR black hole should never be a solution with higher mass/energy.

A first hint about dynamical stability may be obtained by considering the thermodynamic stability of solutions, i.e., by comparing the entropy of curvaturized black holes and Schwarzschild black holes and assuming that dynamics will favor the solution with larger entropy. To compute the entropy, it is useful to introduce the area of the event horizon, which
is given by $A_H = 4\pi r_H^2$. The entropy of the black hole is then given by \cite{Wald:1993nt}
\begin{equation}
	S = \frac{A_H}{4}\R_H + 4\pi \xi(\R_H).\label{eq:entropy}
\end{equation}
The first term shows a standard scaling with the area that one expects for black holes based, among others, on holographic ideas. The second term depends on $\R_H$, i.e., the curvature evaluated at the horizon, but does not manifestly show a scaling with area. We can, however, provide an estimate for this term by using the inequality \eqref{eq:existence_conditions}. We assume that we can saturate the inequality and integrate it locally to obtain $\xi(\R_H) = \frac{r_H^2}{8}\R_H + \rm const$. Plugging this into Eq.~\eqref{eq:entropy}, we find that the second term also exhibits an area scaling, with a smaller coefficient than the first term. We do therefore not expect large deviations of the entropy from the Bekenstein-Hawking entropy for a Schwarzschild black hole. This expectation is confirmed by our numerical results, cf.~Fig.~\ref{fig:plots}.

From the bottom-left plot in Fig.~\ref{fig:plots}, we observe that the entropy of the curvaturized black holes is smaller than that of a GR black hole when $M/\sqrt{\alpha} > 1.174$ (i.e., when $\R_H>1$), such that these solutions (indicated in black in Fig.~\ref{fig:plots}) are not entropically preferred.\footnote{Our results do not confirm the study in \cite{Liu:2020yqa}, where entropy and Hawking temperature of the curvaturized solution was found to agree with those of the Schwarzschild black hole of the same ADM mass.} In many theories, there is a link between thermodynamic instability and dynamical instability; there are, however, counterexamples, see, e.g., \cite{Held:2022abx}. 

Solutions with $\R_H<1$ (indicated in green in Fig.~\ref{fig:plots}) are generally entropically preferred with respect to their GR counterparts and constitute a minimum-mass solution, beyond which curvaturized black holes no longer exist. The minimum-mass solution is non-singular on and outside the horizon. 

However, there is another branch of solutions (indicated in red in Fig.~\ref{fig:plots}) with values of $\R_H$ smaller than those of the minimum-mass solution, but with a higher mass. For this branch entropic preference quickly changes, and these solutions soon present entropies smaller than those of GR and other curvaturized black holes. This second branch of curvaturized solutions is expected to be unstable, as was observed in related theories such as dilaton-Gauss-Bonnet gravity \cite{Torii:1996yi,Corelli:2022phw}, and the endpoint of the branch is singular -- the horizon overlaps with a curvature singularity once the conditions in Eq.~\eqref{eq:existence_conditions} are saturated, much like in the examples studied in Ref.~\cite{Fernandes:2022kvg}.

The Hawking temperature \cite{Hawking:1975vcx} of the curvaturized black holes is
\begin{equation}
	T = \frac{1}{4\pi} f'_H e^{-\delta_H}.
\end{equation}
Curvaturized black holes are generally found to have a Hawking temperature $T$ higher than that of Schwarzschild black holes, irrespective whether their entropy is smaller or larger than that of their GR counterparts. 
Both the entropy and Hawking temperature are of course frame invariant \cite{Jacobson:1993vj}, i.e., they yield the same value in both the Jordan and Einstein frames, as they should as observables.

As shown in the top-right plot of Fig.~\ref{fig:plots}, the scalaron charge $Q$ follows a profile similar to that of the scalaron at the horizon and has the same sign as $\R_H - 1$.

\begin{figure*}[]
	\centering
	\includegraphics[width=\linewidth]{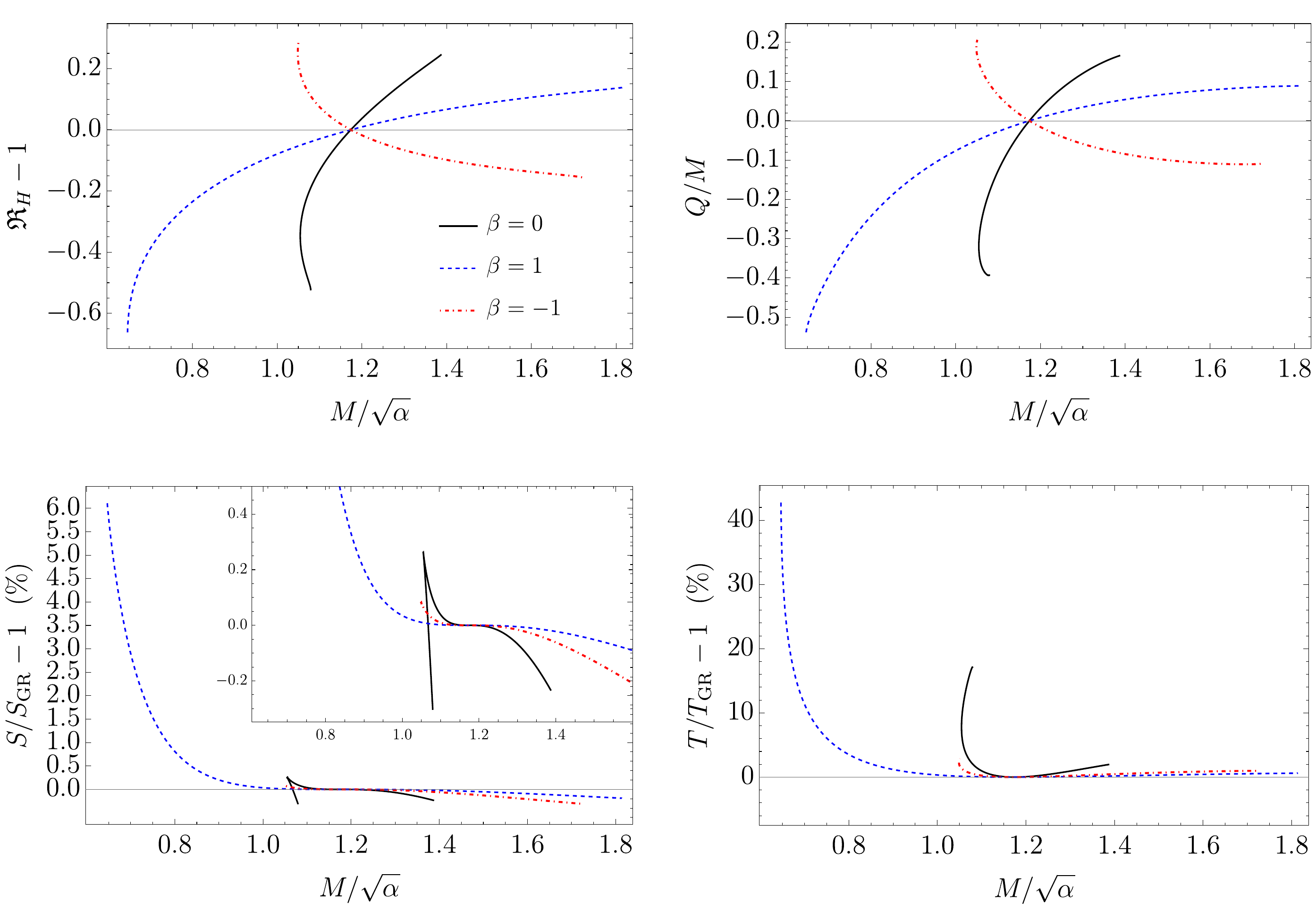}\vfill
    \caption{Domain of existence of curvaturized black holes with a Ricci-type coupling and their physical quantities. The values of $\R_H$ (top-left), $Q$ (top-right), $S$ (bottom-left) and $T$ (bottom-right) of curvaturized black holes are presented as a function of $M/\sqrt{\alpha}$. The case $\beta=1$ is presented in the dashed blue line, while $\beta=-1$ is shown in the red dotted line.}
	\label{fig:plotsbeta}
\end{figure*}

\subsection{Curvaturized black holes with a non-zero Ricci-coupling}
\label{sec:ricci-coupling}

Although the tachyonic instability of Schwarzschild black holes is linear, meaning only quadratic interactions of the scalaron contribute to it, the instability is quenched by the system’s non-linearities. As a result, scalarized black holes in the canonical scalarization scenario with a quadratic coupling \cite{silvaSpontaneousScalarizationBlack2018} are unstable, requiring additional non-linear interactions to stabilize them \cite{Macedo:2019sem,antoniouStableSpontaneouslyscalarizedBlack2022}. This implies that additional couplings, which may not affect the theory when linearized around a GR black hole, play a crucial role in determining the properties of scalarized black holes. A notable example is a coupling to the Ricci scalar \cite{antoniouStableSpontaneouslyscalarizedBlack2022,ventagliNeutronStarScalarization2021,Fernandes:2024ztk,antoniouBlackHoleScalarization2021,thaalbaSphericalCollapseScalarGaussBonnet2023,Thaalba:2024htc}.
Ricci-type couplings have been demonstrated to stabilize static and spherically symmetric scalarized black holes \cite{antoniouStableSpontaneouslyscalarizedBlack2022,antoniouBlackHoleScalarization2021}, (see, however, \cite{Kleihaus:2023zzs}) enable scalarization models to bypass binary pulsar constraints \cite{ventagliNeutronStarScalarization2021}, and alleviate the loss of hyperbolicity in these theories \cite{thaalbaSphericalCollapseScalarGaussBonnet2023,Thaalba:2024htc}, among other benefits \cite{Antoniou:2020nax}.

While we have not investigated stability properties of curvaturized black holes explicitly, we expect that our theory is affected by the same issues as standard curvaturization scenarios without Ricci couplings. We thus focus on the theory in Eq.~\eqref{eq:scalarization2} at nonzero $\beta$.
Specifically, we examine the cases $\beta = \pm 1$ and compare them to the case discussed in the previous section, where the Ricci-type coupling is absent ($\beta = 0$).
The results are shown in Fig.~\ref{fig:plotsbeta}, where we present the values of $\R_H$ (top-left), $Q$ (top-right), $S$ (bottom-left), and $T$ (bottom-right) as functions of $M/\sqrt{\alpha}$. 

With a non-zero Ricci-type interaction, the domain of existence extends over a broader range of $M/\sqrt{\alpha}$ values for both positive and negative $\beta$. In both cases, the minimum-mass solution marks has a different status than in the theory at $\beta=0$. Whereas it previously connected two branches of the curvaturized solution, one of which had a singular endpoint, it now constitutes the singular endpoint of the curvaturized branch
and saturates the conditions in Eq.~\eqref{eq:existence_conditions}. This is due to the fact that the curve that constitutes the domain of existence, rotates in the $\{\R_H-1,M/\sqrt{\alpha}\}$-plane. Thus, there is a critical value of $\beta$ (both for negative and positive $\beta$), at which the minimum-mass solution becomes the endpoint of this curve.

The general trend regarding entropic preference and Hawking temperature remains the same as in the $\beta=0$ case, where solutions with $M/\sqrt{\alpha} \lesssim 1.174$ are entropically preferred, and generally all solutions exhibit a higher Hawking temperature. However, for sufficiently negative values of $\beta$, the entropically preferred region of the curvaturized branch occurs when $\R_H > 1$, in contrast to the vanishing and positive cases of $\beta$. 

Deviations from GR can become pronounced for $\beta=1$. We leave a study of observationally excluded values of $\alpha$ and $\beta$ to future work.

\subsection{Rotating curvaturized black holes}
\label{sec:rotating}
There is evidence that astrophysical black holes have spin, both in the LVK mass range, see \cite{KAGRA:2021duu}, as well as in the range of supermassive black holes \cite{EventHorizonTelescope:2019pgp,Reynolds:2020jwt}. It is therefore crucial to go beyond spherical symmetry and confirm whether curvaturized solutions exist at nonzero spin.

To obtain the rotating curvaturized solutions, we use the code and follow closely the numerical method of Ref.~\cite{Fernandes:2022gde}. We consider a circular metric ansatz in quasi-isotropic coordinates with four functions $f,g,h,W$ of $r$ and $\theta$
\begin{equation}
\begin{aligned}
    ds^2 =& -f \left(1-\frac{r_H}{r}\right)^2 dt^2 + \frac{g h}{f} \left(dr^2 + r^2 d\theta^2\right),\\& + \frac{g}{f} r^2 \sin^2\theta \left(d\varphi - \frac{W r_H}{r^2} dt\right)^2,
\end{aligned}
\label{eq:metric}
\end{equation}
where $r_H$ is the coordinate location of the event horizon. The radial coordinate is compactified as $x=1-2r_H/r$, mapping the interval $[r_H, +\infty)$ to $[-1, 1]$. All functions have definite (even) parity with respect to $\theta=\pi/2$, thus we consider only the range $\theta \in [0, \pi/2]$ in our numerical setup.

The four metric functions and the scalar field are subject to boundary conditions. Regularity, axial symmetry and parity considerations imply $\partial_\theta f = \partial_\theta g = \partial_\theta h = \partial_\theta W = \partial_\theta \R = 0$, at $\theta = 0, \, \pi/2$. At the horizon, $x=-1$, the functions obey $f-2 \partial_x f = g+2 \partial_x g = \partial_x h = W - r_H \Omega_H = \partial_x \R = 0$, where $\Omega_H$ is the angular velocity of the horizon. Asymptotic flatness results in boundary conditions at $x=1$, namely $f=g=h=\R=1$, and $W = 0$. The angular momentum $J$ can be extracted from the asymptotic fall-off of the metric function $g_{\varphi t} \sim 2J\sin^2 \theta/r^2 + \mathcal{O}\left(r^{-3}\right)$. We define the dimensionless spin $j\equiv J/M^2$.

To solve the partial differential field equations, we have used the code described in Ref.~\cite{Fernandes:2022gde}, which employs a pseudospectral method together with the Newton-Raphson root-finding algorithm \cite{numericalMethodsDias}. Each of the functions is expanded in a spectral series with resolution $N_x$ and $N_\theta$ in the radial and angular coordinates $x$ and $\theta$, respectively. The spectral series used for each of the functions $\mathcal{F}^{(k)}=\{f,g,h,W,\R\}$ is
\begin{equation}
  \mathcal{F}^{(k)} = \sum_{n=0}^{N_x-1} \sum_{m=0}^{N_\theta-1} c_{nm}^{(k)} T_n(x) \cos \left(2m\theta\right),
\label{eq:spectralexpansion1}
\end{equation}
where $T_n(x)$ denotes the $n^{\rm th}$ Chebyshev polynomial, and $c_{nm}^{(k)}$ are the spectral coefficients. The angular boundary conditions are automatically satisfied with this spectral expansion, and need not be explicitly imposed in the numerical method.

There are three input parameters: $(r_H, \Omega_H, \alpha)$. We have observed exponential convergence with an increase of resolution, and found that a resolution of $N_x \times N_\theta = 40 \times 8$ is high enough to obtain an estimated error of at most $\mathcal{O}\left(10^{-8}\right)$ using a Smarr-type relation derived as in Ref.~\cite{Fernandes:2022gde}.

\begin{figure}[]
    \centering
    \includegraphics[width=\linewidth]{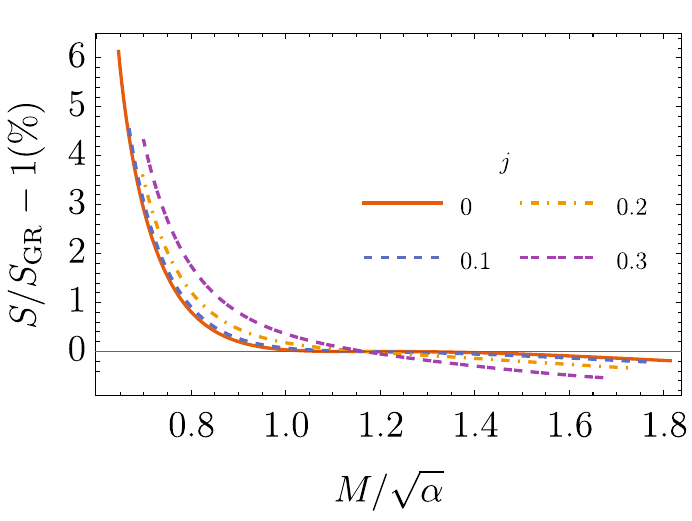}\vfill
    \includegraphics[width=\linewidth]{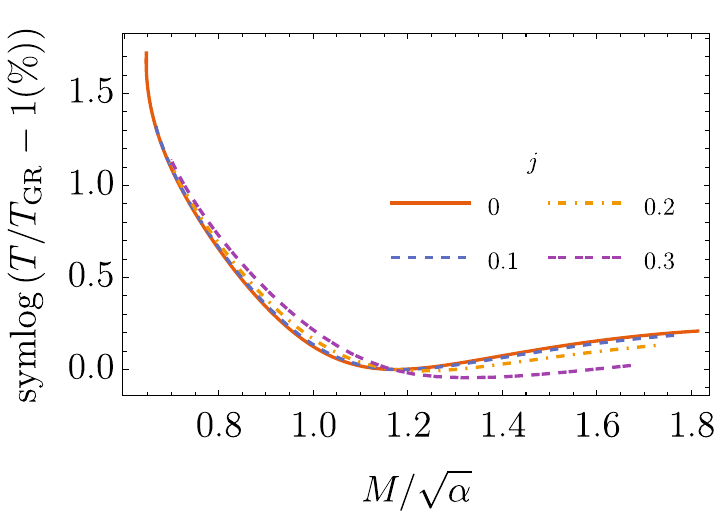}\vfill
    \caption{Entropy (left) and temperature (right) of rotating black holes of the theory \eqref{eq:scalarization2} with a Ricci-type coupling $\beta=1$. We have defined $\mathrm{symlog}(x) = \mathrm{sgn}(x) \log_{10}\left( |x|+1 \right)$. The results are presented for four distinct values of dimensionless spins $j=J/M^2$, where $J$ is the black hole's angular momentum.}
    \label{fig:plotsrotation}
\end{figure}

We find rotating curvaturized black holes that continuously extend the spherically symmetric solution to finite spin, see Fig.~\ref{fig:plotsrotation}. We show results for the entropy and Hawking temperature of rotating curvaturized black holes with spins between $j=0$ and $j=0.3$ of the theory \eqref{eq:genact1} with $\beta=1$. The properties of the rotating black holes qualitatively follow the same trend as their static counterparts, and thus our conclusions from section \ref{sec:ricci-coupling} hold.  At a fixed value of $M/\sqrt{\alpha}$, the deviation of curvaturized rotating black holes from Kerr black holes increases with increasing spin. We also see that the domain of existence shrinks as the spin increased, as expected from the literature on scalarized black holes \cite{cunhaSpontaneouslyScalarisedKerr2019}. From a phenomenological perspective, this is interesting, because it implies that curvaturization is easier to detect at larger spins, but the target range of masses shrinks.

\section{Conclusions and outlook}
\label{sec:conclusions}
GR does not provide a complete description of black holes and must therefore be modified, e.g., by quantum effects. Both classical modifications of GR as well as quantum modifications can be encoded in an effective action. Ultimately, the goal of quantum gravity research is to derive such an effective action, but this goal has only partially been reached. Against this background, we take a different approach. We postulate that the effective action should be formulated purely in terms of curvature invariants. The theory should also be free from pathologies; in particular, there should be stable cosmologies (close enough to $\Lambda$CDM to match observations) and stable black-hole solutions (again, close enough to GR solutions to match observations). It is currently not known in full generality what conditions are necessary or sufficient to guarantee a theory's stability. To motivate a class of actions, we therefore proceed as follows: first, we discard Riemann and Ricci tensor curvature invariants, because in local theories these are generically tied to massive, ghostlike spin-2 degrees of freedom.\footnote{Whether these ghosts are truly catastrophic is subject of an ongoing debate, both at the classical \cite{Held:2023aap} and quantum level \cite{Donoghue:2019ecz}.} This leaves us with $f(R, \G)$ theories, which, however, generically feature instabilities as well when they propagate two scalars. We therefore focus on $f(R, \G)$ theories which satisfy a degeneracy condition and thus only propagate one scalar degree of freedom.
These theories are equivalent to a subset of Horndeski theories, which we identify explicitly. 

Next, we choose a more specific subclass of $f(R, \G)$ theories, namely those that have two black-hole branches, a Kerr branch as well as a new branch for which the Ricci scalar is non-zero. This is motivated by looking for a pathway to test a subclass of the proposed modified theories of gravity, e.g., through black-hole observations.
We construct this subclass by exploiting the equivalence to Horndeski theories, in which the conditions for black-hole scalarization are well-established.  Based on this equivalence, we find the model introduced in Ref.~\cite{Liu:2020yqa}, and generalizations thereof.
In our case, because the scalar is not an independent scalar field, but part of the metric, we call the analogous mechanism \emph{black-hole curvaturization}. According to this mechanism, sufficiently small black holes experience a tachyonic instability driven by the spacetime curvature. This model serves as a proof-of-concept for how spontaneous scalarization-type phenomena can occur without explicitly introducing a scalar field, relying solely on interactions involving the metric.

We establish that curvaturized black holes exist and study their domain of existence in the space spanned by two of the couplings that parameterize degenerate $f(R, \G)$ theories. We also go beyond spherical symmetry and study curvaturized solutions at finite spin. We observe that deviations from GR can become more pronounced at nonzero Ricci coupling and increase with spin when the mass of the black-hole is held fixed. A full study of the space of couplings and parameters of the black hole, most importantly its spin, is left to future work, to determine which combination of coupling values and parameters maximizes deviations from GR and which observable shows the largest deviation. 

In addition, we do not explore the spacetime region inside the horizon. It would indeed be interesting to discover whether the spacetime is geodesically incomplete, whether curvature invariants diverge and whether there is an inner horizon (which is generically subject to a mass-inflation instability) -- all problems that afflict Kerr black holes. We expect a finite radius singularity to exist in most cases, as we have observed that curvaturized solution branches end on a critical singular solution with non-zero mass. As studied in Ref.~\cite{Fernandes:2022kvg}, this usually signals the overlap of an inner singularity with the horizon of the black-hole.

While our focus has been on the gravitational sector, a matter sector must also be introduced if the theory is to be relevant for astrophysical black holes. Matter fields can, in principle, couple to either the Jordan frame metric $g_{\mu \nu}$ or to Einstein frame metric $\R g_{\mu \nu}$. For a coupling to the Jordan frame metric, a fifth force may be induced and result in experimental constraints on the theory, unless there is also a screening mechanism present.
In fact, the non-linear interactions present in the Einstein frame (see Appendix \ref{app:EinFrame}) could potentially give rise to Vainshtein-type screening mechanisms. As demonstrated in Ref.~\cite{gannoujiVainshteinMechanismGaussBonnet2012}, a theory derived from the dimensional reduction of higher-dimensional Einstein-Gauss-Bonnet gravity exhibits Vainshtein screening due to the non-linear interactions of the dilaton in the Einstein frame. Their Einstein-frame action shares a very similar structure with ours, suggesting that our theory may also exhibit some form of non-linear screening. 
We leave the question of whether a fifth force arise and if yes, whether it is screened, to future work.

Additionally, it would be interesting to explore whether spin-induced curvaturization \cite{herdeiroSpininducedScalarizedBlack2021,dimaSpininducedBlackHole2020,bertiSpininducedBlackHole2021} is also possible in this model. This is in principle straightforward along the lines of our study in Sec.~\ref{sec:rotating}, but we leave it to a future study aimed at finding where in coupling and parameter space deviations from GR are maximized.

Although we expect curvaturized solutions to be stable, at least in some portion of the parameter space, their stability remains another important open question, see \cite{Kleihaus:2023zzs} for work on the stability of the scalarized case.\\
In addition to the stability of the curvaturized solutions, there is an even more serious question that the closely related scalar-Gauss-Bonnet theories raise: while they admit a well-posed initial value problem, numerical evolutions of the field equations that describe a binary black-hole merger give rise to elliptic regions in the spacetime, where time evolution breaks down \cite{Ripley:2019irj,Ripley:2020vpk,East:2020hgw,East:2021bqk,Corman:2022xqg}. This brings us back to our starting point, namely the quest for well-defined modified gravity theories beyond GR that are good candidates for the effective action of gravity. We consider degenerate $f(R,\G)$ theories that exhibit curvaturization particularly interesting
candidates for numerical studies, first, to explore how much gravitational waveforms deviate from their GR counterparts, and second, to understand whether the constructed theories are well-behaved in the fully dynamical regime.

The degeneracy condition $\left(\partial_R^2 f \right)\cdot \left(\partial_\G^2 f \right)-\left(\partial_R \partial_\G f\right)^2=0$ must be satisfied to avoid a second propagating scalar. It is an intriguing question how this condition evolves under the Renormalization Group (RG) flow, when quantum fluctuations of gravity are accounted for. Concretely, for our purposes, it is most important to know whether perturbations $\delta f(R, \G)$ are irrelevant under the RG flow and therefore decrease, when quantum fluctuations are integrated out. In case it is irrelevant, degenerate $f(R, \G)$ theories serve as an attractor under the RG flow and do indeed constitute candidates for the effective action of quantum gravity.

%%%%%%%%%%%%%%%%%%%%%%%%%%%%%%%
\section*{Acknowledgments}
%%%%%%%%%%%%%%%%%%%%%%%%%%%%%%%
\noindent A.~E.~and P.~F.~acknowledge support by a grant from Villum Fonden under Grant No.~29405. This work is funded by the Deutsche Forschungsgemeinschaft (DFG, German Research Foundation) under Germany’s Excellence Strategy EXC 2181/1 - 390900948 (the Heidelberg STRUCTURES Excellence Cluster).

\appendix

\section{The Einstein Frame}
\label{app:EinFrame}
To express the set of theories in Eq.~\eqref{eq:puregravity} in the Einstein frame, we can perform the conformal transformation given in Eq.~\eqref{eq:conformaltransf}. The metric determinant, Ricci scalar and Gauss-Bonnet invariant transform as
\begin{equation}
	\begin{aligned}
		&\sqrt{-g} = e^{-2\sqrt{2/3} \varphi} \sqrt{-\tilde{g}},\\&
		\sqrt{-g} R = \sqrt{-\tilde{g}} e^{-\sqrt{2/3} \varphi} \left[ \tilde{R} + \sqrt{6} \tilde{\Box} \varphi - (\partial \varphi)^2 \right],\\&
		\sqrt{-g} \G = \sqrt{-\tilde{g}} \left[ \tilde{\G} + \tilde{\nabla}_\mu J^\mu \right],
		\label{eq:conformal1}
	\end{aligned}
\end{equation}
where
\begin{equation}
	\begin{aligned}
		J^\mu =& \frac{4}{3}\bigg[\partial^\mu \varphi \tilde{\Box} \varphi -\sqrt{6}\tilde{G}^{\mu \nu} \partial_\nu \varphi \\& -  \tilde{\nabla}^\mu \partial^\nu \varphi \partial_\nu \varphi - \frac{1}{\sqrt{6}} \partial^\mu \varphi (\partial \varphi)^2 \bigg].
	\end{aligned}
\end{equation}
Then, in the Einstein frame, the theory in Eq.~\eqref{eq:puregravity} becomes
\begin{equation}
	\begin{aligned}
		S =\int \D^4x \sqrt{-\tilde{g}} \bigg[& \tilde{R} - (\partial \varphi)^2 - e^{-2\sqrt{2/3} \varphi} V\left(e^{\sqrt{2/3} \varphi}\right)\\&
		+ \xi\left(e^{\sqrt{2/3} \varphi}\right) \big( \tilde{\G} + \tilde{\nabla}_\mu J^\mu \big) \bigg],
	\end{aligned}
	\label{eq:einsteinframe}
\end{equation}
where we have taken $\R = e^{\sqrt{2/3} \varphi}$. It becomes manifest that in the Einstein frame we have a scalar-tensor theory with a canonical kinetic term, a scalar potential, a scalar-Gauss-Bonnet coupling and non-trivial higher-order interaction terms. The theory \eqref{eq:einsteinframe} presents the most general Horndeski theory in the Einstein frame that can be expressed in a pure-gravity formulation. It is the hidden non-trivial structure of the terms arising from the conformal transformation of $\G$ that differentiate this special class of theories from more standard scalar-Gauss-Bonnet models \cite{donevaScalarization2022}. Upon integration by parts, and discarding total derivatives, we find that the Einstein frame's theory \eqref{eq:einsteinframe} is equivalent to
\begin{equation}
	\begin{aligned}
		S = & \int \D^4x \sqrt{-\tilde{g}} \bigg[ \tilde{R} - (\partial \varphi)^2 - e^{-2\sqrt{2/3} \varphi} V \\&
		+ \xi \,\tilde{\G} + e^{\sqrt{2/3} \varphi} \bigg(\frac{8}{3} \tilde{G}^{\mu \nu} \partial_\mu \varphi \partial_\nu \varphi \, \xi'\\&
		-2\sqrt{\frac{2}{3}} \tilde{\Box} \varphi (\partial \varphi)^2 \xi' - \frac{4}{9} (\partial \varphi)^4 e^{\sqrt{2/3} \varphi} \xi'' \bigg) \bigg],
	\end{aligned}
	\label{eq:einsteinframe2}
\end{equation}
where the primes denote differentiation with respect to the argument $\R=e^{\sqrt{2/3} \varphi}$.

\bibliography{biblio}

\end{document}